\documentclass[aps,pra,twocolumn,superscriptaddress,nofootinbib]{revtex4}
\usepackage{mathptmx}
\usepackage{amsmath,amssymb, amscd}
\usepackage[pdftex]{graphicx}
\usepackage{hyperref}
\usepackage{bm}

\newcommand{\be}{\begin{equation}}
\newcommand{\ee}{\end{equation}}

\begin{document}

\title{Quantum-state steering in optomechanical devices}

\author{Helge M\"uller-Ebhardt}
\affiliation{Max-Planck-Institut f\"ur Gravitationsphysik
(Albert-Einstein-Institut) and Universit\"at Hannover, Callinstr.~38, 30167 Hannover, Germany}
\author{Haixing Miao}
\affiliation{California Institute of Technology, MC 350-17,
Pasadena, California 91125}
\author{Stefan Danilishin}
\affiliation{School of Physics, University of Western Australia, WA 6009, Australia}
\author{Yanbei Chen}
\affiliation{California Institute of Technology, MC 350-17,
Pasadena, California 91125}


\begin{abstract}
We show that optomechanical systems in the quantum regime can be used to demonstrate EPR-type quantum entanglement between the optical field and the mechanical oscillator, via quantum-state steering.  Namely, the conditional quantum state of the mechanical oscillator can be steered into different quantum states depending the choice made on which quadrature of the out-going field is to be measured via homodyne detection.
More specifically, if quantum radiation pressure force dominates over thermal force, the oscillator's quantum state is steerable with a photodetection efficiency as low as 50\%, approaching the ideal limit shown by Wiseman and Gambetta [Phys. Rev. Lett. {\bf 108}, 220402 (2012)].
We also show that requirement for steerability is the same as those for achieving sub-Heisenberg state tomography using the same experimental setup.
\end{abstract}


\maketitle

{\it Introduction.}---Optomechanical devices are now approaching the quantum regime
and can therefore provide a platform for investigating quantum behaviors of macroscopic mechanical
degrees of freedom\,\cite{Aspelmeyer2012}.
 A topic extensively discussed is how to demonstrate quantum
entanglement in such systems\,\cite{Mancini2002, Pinard2005, Vitali2007, Paternostro2007, Mueller-Ebhardt2008, Hartmann2008, Miao2010b}. In this paper, we focus on an interesting aspect
of bipartite quantum entanglement---the ability of modifying the quantum state
of one party by making different measurements on the other.
This is essence of the Gedankenexperiment of Einstein, Podolsky, and Rosen (EPR)\,\cite{Einstein1935}, and has been rigorously formulated by Wiseman {\it et al.}\,\cite{Wiseman2007,Jones2007,Saunders2010} as quantum steerability.
More recently, Wiseman has shown that steerability can be demonstrated by showing detector-dependent stochastic evolution of a two-level atom coupled to an optical field which in turn is measured continuously\,\cite{Wiseman2012}.

\begin{figure}[!b]
\includegraphics[width=0.45\textwidth]{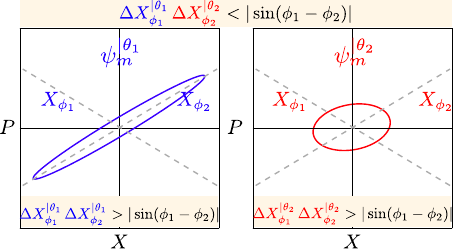}
\caption{(color online) Two different quantum states (projection of their
Wigner functions on phase space) of a mechanical
oscillator, conditional on two different strategies for measuring the optical quadrature: one at $\theta_1(t)$
(left) and the other at $\theta_2(t)$  (right).
\label{steering}}
\end{figure}

In the context of linear quantum systems with Gaussian states (appropriate for optomechanics experiments),   steering can be understood as follows. According
to quantum mechanics, position and momentum of a mechanical oscillator satisfy the Heisenberg uncertainty principle, which reads:
\be
\Delta X_{\phi_1} \, \Delta X_{\phi_2} \ge |\sin(\phi_1-\phi_2)|\,, \quad \forall \phi_1,\phi_2\,
\ee
where $\hat X_{\phi}\equiv (\hat x/\Delta x_q)\sin\phi+(\hat p/\Delta p_q)\cos\phi$, with $\Delta  x_q$ and $\Delta p_q$
zero-point uncertainties in position and momentum,  are quadratures of the mechanical oscillator. Let us assume a mechanical oscillator
is interacting and establishing entanglement with a continuous optical field, and then the field gets continuously measured via time-dependent homodyne detection with $\theta(t)$ being the  optical quadrature measured at time $t$. Suppose the measurement lasts from $-\tau$ up to $0$, the final conditional state of the oscillator, written as $|\psi_m^{|\theta}(0)\rangle$, will depend on how we make the homodyne detection due to entanglement. For two different measurement strategies with $\theta_1(t)$ and $\theta_2(t)$, respectively, in general we have two different final conditional states: $|\psi_m^{|\theta_1}(0)\rangle\neq
|\psi_m^{|\theta_2}(0)\rangle$. If the quadratures are properly chosen, we may have, as illustrated in Fig.\,\ref{steering}:
\begin{equation}\label{ineq}
\Delta X_{\phi_1}^{|\theta_1} \, \Delta X_{\phi_2}^{|\theta_2}  < |\sin(\phi_1-\phi_2)|\,,
\end{equation}
where $\Delta X_{\phi_k}^{|\theta_k}\equiv
\langle(\hat X_{\phi_k}-\langle \hat {X}_{\phi_k}\rangle)^2\rangle^{1/2}$ with $\langle \cdot \rangle \equiv \langle  \psi_m^{|\theta_k}|\cdot |  \psi_m^{|\theta_k}\rangle$.
In other words, if in the first strategy, the observer tries to predict quadrature $ X_{\phi_1}$ of the mechanical oscillator, while in the second strategy, the observer tries to predict $X_{\phi_2}$, then the two predictions have an error product that is {\it lower} than Heisenberg Uncertainty.  This is the essential feature of the EPR paradox; the ability of an experimental setup in creating such a feature is called {\it quantum steerability}~\cite{Wiseman2007,Jones2007,Saunders2010}.

In ideal linear quantum measurement processes, both conditional states will be pure, and for any pairs of distinctive $\theta_1$ and $\theta_2$, inequality~\eqref{ineq} will almost always exist for some set of $\phi_1$ and $\phi_2$---although this idealized steerability may be influenced by practical imperfections, such as thermal noise.  In view of Eq.\eqref{ineq}, we introduce a figure of merit to
quantify steerability,
\be\label{steerability}
{\cal S}\equiv - \min_{\phi_1,\phi_2,\theta_1,\theta_2}\left\{\ln\frac{\Delta X_{\phi_1}^{|\theta_1}\,
\Delta X_{\phi_2}^{|\theta_2}}{|\sin(\phi_1-\phi_2)|}\right\}.
\ee
with minimum obtained by
comparing all possible sets of $\{\phi_1, \theta_1(t),  \phi_2, \theta_2(t)\}\; (t\in[-\tau,\,0])$---an optimal time-dependent homodyne detection is needed to achieve the lower bound.
The quantum state is steerable when ${\cal S}>0$, which will be proved to be equivalent to the formal criterion obtained by Wiseman for Gaussian entangled states\,\cite{Wiseman2007}.
As we will show in the discussion that follows, for linear optomechanical devices, when the quantum radiation pressure dominates strongly over
 thermal fluctuations, steerability only depends on
the photodetector efficiency $\eta$ of time-dependent homodyne detections:
\be\label{eq_S}
{\cal S} \approx \frac{1}{2}\left[\ln\eta-\ln(1-\eta)\right],
\ee
which will be positive as long as $\eta > 50\%$, which coincides with the ideal limit shown by Wiseman and Gambetta\,\cite{Wiseman2012}.
Interestingly, such quantum steerability is intimately related to the state tomography accuracy in the
protocol suggested by Miao {\it et al.}~\cite{Miao2010}, in which an optimal time-dependent homodyne
detection scheme is used to probe the quantum state of a mechanical oscillator with Gaussian-distributed
joint position and momentum error
less than Heisenberg uncertainty. More explicitly, we will show, for the same optomechanical device,
\be
{\cal S} = -\ln\left[{2}\sqrt{\det{\bf V}_v}/\hbar\right],\,
\ee
where ${\bf V}_v$ is the covariance matrix for the tomograph error.

{\it Optomechanical dynamics.---}We start by considering dynamics of linear optomechanical device which
has been extensively studied\,\cite{Marquardt2007, Wilson-Rae2007, Genes2008a, Miao2010b}; its
linearized Hamiltonian reads:
\begin{align}\nonumber
\hat {\cal H}=&{\hat p^2}/{(2m)}+m\omega_m^2 \hat x^2/2 +\hat {\cal H}_{\kappa_m}+\hbar\, \Delta
\hat a^{\dag}\hat a+\hbar g\hat x(\hat a^{\dag}+\hat a)\\&+i\hbar\sqrt{\kappa}
(\hat a_{\rm ext}\hat a^{\dag}-\hat a_{\rm ext}^{\dag}\hat a)\,.
\label{hamiltonian}
\end{align}
Here $\omega_m$ is the mechanical resonant frequency; $\Delta=\omega_c-\omega_l$ is cavity detuning,
i.e., difference between the cavity resonant frequency $\omega_c$ and the laser frequency $\omega_l$;
$\hat {\cal H}_{\kappa_m}$ summarizes the fluctuation-dissipation mechanism for the mechanical
oscillator; the fifth term is the optomechanical interaction term with $g\equiv \bar a\,\omega_c/L$
quantifying the coupling strength, $\bar a$ the steady-state amplitude of the cavity mode and $L$
the cavity length; the last term describes the coupling between the cavity mode and external continuous
optical field $\hat a_{\rm ext}$ with $[\hat a_{\rm ext}(t),\,\hat a_{\rm ext}^{\dag}(t')]=\delta(t-t')$
in the Makovian limit and $\kappa$ is the coupling rate which is also the cavity bandwidth.

From Hamiltonian~\eqref{hamiltonian}, one can obtain the following set of linear Heisenberg equations of motion
\begin{align}\label{eom1}
m[\ddot{\hat x}(t)+\kappa_m\dot{\hat x}(t)+\omega_m^2\hat x(t)] &=\hat F_{\rm rp}(t)+\hat F_{\rm th}(t)\,,\\
\label{eom2}
\dot{\hat a}(t)+({\kappa}/{2}+i\,\Delta)\hat a(t)&=-ig\hat x(t)+\sqrt{\kappa}\,\hat a_{\rm in}(t)\,,
\end{align}
 and the input-output relation
\be\label{eom3}
\hat a_{\rm out}(t)=-\hat a_{\rm in}(t)+\sqrt{\kappa}\,\hat a(t)\,,
\ee
where $\hat a_{\rm in}\equiv \hat a_{\rm ext}(t_-)$ (in-going) and $\hat a_{\rm out}\equiv \hat a_{\rm ext}(t_+)$ (out-going) are input and output operators in the standard input-output formalism\,\cite{Walls2008}, and $\hat F_{\rm rp}\equiv -\hbar g(\hat a+\hat a^{\dag})$ is the
quantum radiation pressure force and $\hat F_{\rm th}$ is the thermal fluctuation force associated
with the mechanical damping. Output field $\hat a_{\rm out}$ can be measured with a homodyne detection scheme, from which
we can infer the mechanical motion and thus the quantum state of the oscillator. By adjusting the local oscillator phase, one can measure any $\theta$-quadrature: $
\hat b_{\theta}=\hat b_1\sin \theta +\hat b_2\cos\theta$,
which is a linear combination of the output amplitude quadrature $\hat b_1 \equiv \frac{1}{\sqrt{2}} (\hat a_{\rm out}
+\hat a_{\rm out}^{\dag})$
and phase quadrature $\hat b_2\equiv\frac{1}{\sqrt{2}i}(\hat a_{\rm out}-\hat a_{\rm out}^{\dag})$. After incorporating non-unity photodetection efficiency $\eta$, the measurement output at time $t$ is given by
\be\label{eqio}
\hat y_{\theta}(t)=\sqrt{\eta}\left[\hat b_1(t)\sin \theta +\hat b_2(t)\cos\theta\right] + \sqrt{1-\eta}\,\hat n_{\theta}(t)\,,
\ee
where $\hat n_{\theta}$ is the vacuum noise associated with the photodetection loss and is
uncorrelated with $\hat a_{\rm in}$. Note that $\theta$ can be a function of time, when the local oscillator phase is adjusted during the measurement; in this way a different optical quadrature (but only one) is measured at each moment of time.

{\it Conditional quantum state.}---Suppose we perform out homodyne detection during $-\tau \le t \le 0$, obtaining a data string:
\be\label{eqyt}
{\bm y}_{\theta}=\{y_{\theta}(-\tau),\, y_{\theta}(-\tau+{\rm d}t), \,
\cdots, \,y_{\theta}(-{\rm d}t),\,y_{\theta}(0)\}
\ee with ${\rm d}t=\tau/(N-1)$ being the time increment and $N$ the number of data points.
We can then infer the quantum state of mechanical oscillator at $t=0$ conditional
on these measurement data, obtaining the so-called conditional quantum state. The standard way to obtain the conditional state is to use data to  drive
the stochastic master equation\,\cite{{Barchielli1993, Milburn1996, Hopkins2003, Gardiner2004, Doherty1999, Doherty1999a}}.  Here we use a different approach by using the Wigner quasi-probability distributions, deriving the conditional quantum state in a way similar to classical Bayesian statistics, in the same spirit as the approach applied in Ref.\,\cite{Mueller-Ebhardt2009}; this allows us to more straightforwardly treat non-Markovianity, e.g., due to a small cavity bandwidth and colored classical noises. Our approach takes the advantage of the following facts:
\be
[\hat y_{\theta}(t),\,\hat y_{\theta}(t')] = [\hat x(0),\,\hat y_{\theta}(t)]=[\hat p(0),\,\hat y_{\theta}(t)]= 0
\ee
$\forall\, t,t'\in[-\tau, 0]$, which is a consequence of the general features of  linear continuous
quantum measurements\,\cite{BK92}.
We can therefore treat $\hat y_{\theta}(t)$ almost as classical
quantities and ignore their time-ordering in deriving the following joint Wigner function of the oscillator
and the continuous optical field:
\be
W({\bm x}, {\bm y}_{\theta})={\rm Tr}[\hat \rho(-\tau)\delta^{(2)}(\hat {\bm x}-{\bm x})
\delta^{(N)}(\hat {\bm y}_{\theta}- {\bm y}_{\theta})]\,.
\ee
Here we have only included the marginal distribution for the optical quadrature $\hat {\bm y}_{\theta}$ of interests, instead of the entire optical phase space; $\hat {\bm x}\equiv (\hat x(0),\,\hat p(0))$ and $\bm x$ is a c-number vector, similar for ${\bm y}_{\theta}$; $\hat \rho(-\tau)$ is the initial joint density matrix
\be\label{istate}
\hat \rho(-\tau)=\hat \rho_m^{\rm th}\otimes |\bm 0\rangle \langle \bm 0|
\ee
with $\hat \rho_m^{\rm th}$  the thermal state of the oscillator and $|\bm 0\rangle$ the
vacuum state for the optical field---the coherent amplitude of the laser has been absorbed into the
optomechanical coupling constant $g$. Since $\hat x(0)$ and $\hat p(0)$ do not commute we have to explicitly define
$\delta^{(2)}(\hat {\bm x}-{\bm x}) =\int d^2{\bm \xi} e^{-i {\bm \xi} \cdot (\hat{\bm x} -{\bm x})}
$.
Similar to classical Bayesian statistics, the Wigner function for the conditional quantum state of the mechanical
oscillator at $t=0$ reads:
\be\label{bay}
W_m({\bm x}|\,{\bm y}_{\theta})={W({\bm x},{\bm y}_{\theta})}/{W({\bm y}_{\theta})}\,.
\ee
Since we consider only Gaussian quantum states, the joint Wigner function can thus be formally
written as:
\be
W({\bm x}, {\bm y}_{\theta})=c_0 \exp\left[-\frac{1}{2}({\bm x}, {\bm y}_{\theta}){\bf V}^{-1}({\bm x}, {\bm y}_{\theta})^{\rm T}\right]\,,
\ee
where $c_0$ is the normalization factor and superscript $^{\rm T}$ denotes transpose. Elements of the covariance matrix $\bf V$ are given by
\be
{\bf V}_{jk}=\langle \hat {\bm X}_j \hat {\bm X}_k\rangle_{\rm sym}\equiv{\rm Tr}[\hat \rho(-\tau) (\hat {\bm X}_j\hat {\bm X}_k + \hat {\bm X}_k \hat {\bm X}_j)]/2
\ee
with $\hat {\bm X}\equiv (\hat {\bm x},\hat {\bm y}_{\theta})$.
We separate components of the oscillator and the optical field, and
rewrite the covariance matrix $\bf V$ as:
\begin{equation} \label{V}
{\bf V} =\left[\begin{array}{cc}
  {\bf A} & {\bf C}_{\theta}^{\rm T}\\
  {\bf C}_{\theta} & {\bf B}_{\theta} \\
\end{array}\right]\equiv \left[\begin{array}{cc}
{\bf A} & {\bf C}^{\rm T} {\bm u}_{\theta}^{\rm T}\\
{\bm u}_{\theta}{\bf C} & {\bm u}_{\theta} {\bf B}{\bm u}_{\theta}^{\rm T}
\end{array}\right]\,.
\end{equation}
Here $\bf A$ is a $2\times2$ covariance matrix for the mechanical oscillator position $\hat x(0)$ and momentum  $\hat p(0)$;
$\bf B$ is a $2N\times 2N$ covariance matrix for two quadratures $\hat {\bm y}_1\equiv\hat {\bm y}_{\theta =\pi/2}$ and
$\hat {\bm y}_2\equiv \hat {\bm y}_{\theta=0}$ of the optical field; ${\bm u}_{\theta}=(\sin{\bm \theta},\,\cos{\bm \theta})$ is a $N\times 2N$ matrix and $\sin{\bm \theta}\equiv{\rm diag}[\sin\theta(-\tau),\cdots,\sin\theta(0)]$---a diagonal matrix with elements being quadrature
angle at different times; ${\bf C}$ is a $2N\times 2$ matrix describing the correlation between
($\hat {\bm y}_1$,\,$\hat{\bm y}_2$) and ($\hat x(0)$,\, $\hat p(0)$).
Combining Eqs.~\eqref{bay} and \eqref{V}, we obtain:
\be
W_m({\bm x}|\,{\bm y}_{\theta})=\frac{1}{\pi\hbar}\exp\left[-\frac{1}{2}({\bm x}-{\bm x}^{|\theta}){{\bf V}_m^{|\theta}}^{-1}({\bm x}-{\bm x^{|\theta}})^{\rm T}\right]\,,
\ee
where the conditional mean ${\bm x}^{|\theta}$ and covariance matrix  ${\bf V}^{|\theta}_m$ are
\be\label{eq_cond}
{{\bm x}^{|\theta}}=  {\bf C}_{\theta}^{\rm T} {\bf B}_{\theta}^{-1}\bm y_{\theta}^{\rm T}, \quad {\bf V}_m^{|\theta}={\bf A}-{\bf C}^{\rm T}_{\theta}\,{\bf B}_{\theta}^{-1}{\bf C}_{\theta}\,.
\ee
Note that the two rows of the $2N \times 2$ matrix, ${\bf C}^{\rm T}_{\theta}{\bf B}_{\theta}^{-1}$, which we shall refer to as ${\bm K}_x$ (the first row)
and ${\bm K}_p$ (the second row), are also the optimal filters that predict $\hat x(0)$ and $\hat p(0)$ with minimum errors,
$\langle [\hat x(0) - {\bm K}_x \,\hat{\bm y}_{\theta}^{\rm T}]^2 \rangle$ and
$\langle [\hat p(0) - {\bm K}_p \,\hat{\bm y}_{\theta}^{\rm T}]^2 \rangle$, respectively. The above results
for the conditional mean and variance are formally identical to
those obtained by classical optimal filtering.

{\it Quantum-state steering.---}We can now understand the quantum-state steering from a more
quantitative way. From Eq.\,\eqref{eq_cond}, we learned that the conditional variance
of the oscillator state ${\bf V}_m^{|\theta}$ directly depends on the optical quadrature
$\theta$ that we choose to measure. To calculate the steerability figure of merit $\cal S$ [cf. Eq.\,\eqref{steerability}], we need to find the time-dependent quadrature phase $\theta(t)$ that minimize the conditional variance
$\Delta X_{\phi}^{|\theta}$ of a given mechanical quadrature $\hat X_{\phi}={\bm v}_{\phi}\hat {\bm x}^{\rm T}$
with vector ${\bm v}_{\phi}\equiv (\sin\phi/\Delta x_q, \, \cos\phi/\Delta p_q)$. Using the fact that
\begin{align}\nonumber
\min_{\theta}({\Delta X_{\phi}^{|\theta}})^2&=\min_{\theta,\,\bm K}\langle \bm v_{\phi}\hat {\bm x}^{\rm T}-\bm K\hat {\bm y}_{\theta}^{\rm T}\rangle^2\\&
=\min_{\bm K_1,\bm K_2}\langle \bm v_{\phi}\hat {\bm x}^{\rm T}-\bm K_1\hat {\bm y}_1^{\rm T}-\bm K_2\hat {\bm y}_2^{\rm T}\rangle^2\,
\end{align}
with $\bm K_1\equiv \bm K\sin\bm\theta$ and $\bm K_2\equiv \bm K\cos\bm\theta$,  we obtain the minimum
\be\label{eq_DX}
({\Delta X_{\phi}^{|\theta}})^2_{\rm min} =  {\bm v}_{\phi} ({\bf A} - {\bf C}^{\rm T}{\bf B}^{-1}{\bf C}) {\bm v}_{\phi}^{\rm T},
\ee
and $\theta(t_k)$ at $t= -\tau+k\,{\rm d}t$ is given by:
\be\label{eq_theta}
\theta(t_k)=\arctan\left[{{(\bm v}_{\phi}{\bf C}^{\rm T}{\bf B}^{-1})_{k}}/{({\bm v}_{\phi}{\bf C}^{\rm T}{\bf B}^{-1})_{N+k}}\right].
\ee
Since $({\Delta X_{\phi}^{|\theta}})^2_{\rm min}$ is in a quadratic form of ${\bm v}_\phi$, we obtain:
\be\label{eq_steer}
{\cal S}=-\ln\left[{2}\sqrt{\det {\bf V}_s}/\hbar\right],\quad {\bf V}_s\equiv {\bf A} - {\bf C}^{\rm T}{\bf B}^{-1}{\bf C}\,.
\ee
This means quantum state of the oscillator is not steerable---${\cal S}<0$,
if ${\bf V}_s$ is Heisenberg limited---$\sqrt{\det {\bf V}_s} >\hbar/2$.

Such a definition of steerability is in accord with the criterion by Wiseman {\it et al.}\,\cite{Wiseman2007}, more specifically,
shown in their  Eq. (17),  which says that quantum state of the oscillator {\it cannot} be steered by the optical
field, if we have
\be
\left[\begin{array}{cc}
  {\bf A} & {\bf C}^{\rm T} \\
  {\bf C} & {\bf B} \\
\end{array}\right]+i\,{\bm \Sigma}_m\oplus{\bm 0}_o>0\,,
\ee
where $\bm \Sigma_m$ is the $2\times 2$ symplectic matrix for the oscillator, and $\bm 0_o$ is
a null $2N\times 2N$ matrix for the optical field.
Since the covariance matrix ${\bf B}$ for the optical field is positive
definite, namely, ${\bf B}>0$, the above condition requires that the Shur's complement of $\bf A$ be positive
definite:
\be
{\bf A}-{\bf C}^{\rm T}\,{\bf B}^{-1}{\bf C}+i\,{\bm \Sigma}_m={\bf V}_{s}+i\,{\bm \Sigma}_m>0\,,
\ee
which is equivalent to requiring that ${\bf V}_{s}$ is Heisenberg limited, i.e.,
\be
{\cal S}=-\ln\left[ {2}\sqrt{\det{\bf V}_{s}}/{\hbar}\right]
<0\,.
\ee

{\it Continuous-time limit.}---To properly describe the actual continuous measurement process, we take the continuous-time limit with ${\rm d}t\rightarrow 0$, and we have $N\rightarrow \infty$.  The matrices indexed by time become functions of time, while matrix products involving summing over time become integrals.  In particular, the central problem of
calculating ${\bm K} = {\bf C}^{\rm T}{\bf B}^{-1}$ becomes solving an integral equation for ${\bm K}$:
\be\label{eqint}
\int_{-\tau}^0 {\rm d}t'\, {\bf B}(t,\, t') {\bm K}(t')={\bf C}^{\rm T}(t)\,.
\ee
More specifically, ${\bf B}(t,\, t')$ now becomes a $2\times 2$ matrix with elements being the two-time
correlation functions between optical quadratures $\hat y_1(t)$ and $\hat y_2(t')$; ${\bf C}(t)$'s elements are
correlation functions between  $(\hat y_1(t), \hat y_2(t'))$ and $(\hat x(0), \hat p(0))$. These correlation functions can in turn be obtained by solving Heisenberg equations of motion [cf. Eqs.\,(\ref{eom1}-\ref{eqio})]
and expressing $\hat x(0), \,\hat p(0),\, \hat y_1(t)$, and $\hat y_2(t)$
in terms of $\hat a_{\rm in}(t)$, $\hat n(t)$ and $\hat F_{\rm th}(t)$, for which we have
$\langle \hat a_{\rm in}(t)\hat a_{\rm in}^{\dag}(t')\rangle_{\rm sym} =\delta(t-t')/2$,
$\langle \hat n(t)\hat n(t')\rangle_{\rm sym}=\delta(t-t')/2$ and
$\langle \hat F_{\rm th}(t) \hat F_{\rm th}(t') \rangle_{\rm sym}=2m\kappa_mk_B T\delta(t-t')$
given the initial state $\hat \rho(-\tau)$ shown in Eq.\,\eqref{istate}.
The above integral equation is generally difficult to solve analytically if $\tau$ is finite. Since usually we are not interested in the transient dynamics, we can extend
$-\tau$ to $-\infty$, which physically corresponds to waiting long enough till the mechanical oscillator approaches a steady state, and then start state preparation.
In this case, Eq.~\eqref{eqint} can be solved analytical using the Wiener-Hopf method of which the detail is shown in the Appendix\,\ref{App}.

{\it Large cavity bandwidth and strong measurement limit.---}One interesting scenario that allows a nice compact-form
solution is when the cavity bandwidth is large and the optomechanical coupling rate is strong---a strong measurement, compared with the mechanical resonant frequency $\omega_m$.
In this case, the cavity mode can be adiabatically eliminated, and the mechanical resonant frequency $\omega_m$ ignored (for general scenarios, the formalism here still applies but the analytical results become quite complicated). Correspondingly, equations of motion for the oscillator simply
reads [cf. Eqs.\,\eqref{eom1} and \eqref{eom2}]:
\be\label{eq_x}
m\, \ddot{\hat x}(t)=\hat F_{\rm rp}(t)+\hat F_{\rm th}(t)=-\alpha\,\hat a_1(t)+\hat F_{\rm th}(t)\,,
\ee
with $\hat a_1\equiv \frac{1}{\sqrt{2}} (\hat a_{\rm in}+\hat a_{\rm in}^{\dag})$  the amplitude quadrature
of the input field.  The output amplitude quadrature $\hat y_1$ and phase quadrature $\hat y_2$ are given by [cf. Eqs.\,\eqref{eom3} and \eqref{eqio}]:
\begin{align}\label{eq_y1}
\hat y_1(t)&=\sqrt{\eta}\,\hat a_1(t)+\sqrt{1-\eta}\,\hat n_1(t)\,,\\\label{eq_y2}
\hat y_2(t)&=\sqrt{\eta}\,\left[\hat a_2(t)+({\alpha}/{\hbar})\,\hat x(t)\right]+\sqrt{1-\eta}\,\hat n_2(t)\,,
\end{align}
where  $\hat a_2\equiv \frac{1}{\sqrt{2}i}(\hat a_{\rm in}-\hat a_{\rm in}^{\dag})$ is the phase quadrature of
the input field, and additionally we have introduced an effective coupling constant $\alpha\equiv\sqrt{8/\kappa}\,\hbar \,g$. From the above equations, we can easily obtain those correlation functions in the integral equation shown in Eq.\,\eqref{eqint}, solving which leads to:
\be\label{eq_Vs}
{\bf V}_s=\frac{\hbar \zeta_F}{\sqrt{2}\,\eta}\left[\begin{array}{cc}2^{\frac 14}
\sqrt{{ \alpha^2}/{(\zeta_F \hbar m})}&  1\\
1 & 2^{\frac 34}
\sqrt{\zeta_F \hbar m /\alpha^2}
\,
\end{array}\right]\,,
\ee
where the characteristic constant $\zeta_F$ is defined as:
\be
\zeta_F \equiv\left[\frac{{\eta}}{2}\left({1-\eta}+\frac{4m\kappa_mk_B T}{\alpha^2}\right)\right]^{1/2}.
\ee
Correspondingly, we obtain the steerability [cf. Eq.\,\eqref{eq_steer}]:
\be
{\cal S}=-\ln \big({\sqrt{2}\,\zeta_F}/{\eta}\big).
\ee
For a strong measurement, the quantum radiation pressure dominates over the thermal fluctuation force,
and we have
$S_F^{\rm rp}=\alpha^2 \gg S_F^{\rm th}= {4m\kappa_mk_B T}$, with $S_F^{\rm rp}$ and $ S_F^{\rm th}$
being the single-sided spectra density---twice the Fourier transform of two-time correlation function.
This leads to $\zeta_F\approx \sqrt{\eta(1-\eta)/2}$ and $
{\cal S}\approx \frac{1}{2} \ln[{\eta}/({1-\eta})]$, as shown in Eq.\,\eqref{eq_S}.

{\it Connection between steering and quantum tomography.---} Interestingly, such quantum-state steering is closely related to the quantum tomography protocol discussed in Ref.\,\cite{Miao2010}, where an optimal time-dependent homodyne detection is proposed to minimize the error in obtaining marginal distributions of different mechanical quadratures, from which we reconstruct the Wigner function of the quantum state in phase space.
More specifically, for the same optomechanical device discussed above, the tomography error---quantifying the difference between the reconstructed Wigner function and the actual one---is given by the following covariance matrix:
\be\label{eq_Vveri}
{\bf V}_v=\frac{\hbar \zeta_{F}}{\sqrt{2}\,\eta}\left[\begin{array}{cc}2^{\frac 14}
\sqrt{{ \alpha^2}/{(\zeta_{F} \hbar m})}&  -1\\
-1 & 2^{\frac 34}
\sqrt{\zeta_{F} \hbar m /\alpha^2}
\,
\end{array}\right]\,.
\ee
Notice that it is almost identical to the conditional
covariance matrix ${\bf V}_s$ shown in Eq.\,\eqref{eq_Vs}, apart from that the off-diagonal terms have the opposite sign. The state steering can therefore be viewed as the time-reversal counterpart for state tomography, as the off-diagonal term flips sign when the oscillator momentum $\hat p\rightarrow -\hat p$ under $t\rightarrow -t$, and the condition for achieving a sub-Heisenberg error for state tomography---$\sqrt{\det{\bf V}_v}<\hbar/2$, is also identical to that for steerability.

Such a connection can be understood from the fact that for steering, one tries to prepare states with minimal uncertainty in certain quadratures $\hat X_{\phi}(0)$ with data from $(-\infty,\,0]$---a {\it filtering} process, while for tomography, one tries to minimize the error in estimating quadratures $\hat X_{\phi}(0)$ with data from $(0,\,\infty)$---a {\it retrodiction} process. Due to linearity, both the minimal uncertainty and the tomography error for a given quadrature $\hat X_{\phi}$ all takes the quadratic form---${\bm v}_{\phi}\,{\bf V}_{s, \,v}\,{\bm v}_{\phi}^{\rm T}$ and
\be\label{eqSV}
\begin{CD}
{\bf V}_{s}={\bf A}-{\bf C}^{\rm T}{\bf B}^{-1}{\bf C}@>{\mbox{$t\rightarrow -t$}}>>{\bf V}_v\,.
\end{CD}
\ee
These two covariance matrixes ${\bf V}_{s, \,v}$ describe the remaining uncertainty in oscillator
 position $\hat x(0)$ and momentum $\hat p(0)$ conditional
on both the amplitude $\hat y_1$ and phase quadrature $\hat y_2$ of optical field, for $t<0$ and  $t>0$, respectively. Note that the above relation shown in Eq.\,\eqref{eqSV} is exact only when the noise during the state preparation and the one during tomography are uncorrelated, as the correlation between them will break down the time-reversal symmetry, which happens if the cavity bandwidth is small and has noneligible memory time.

\begin{figure}[!t]
\includegraphics[width=0.48\textwidth]{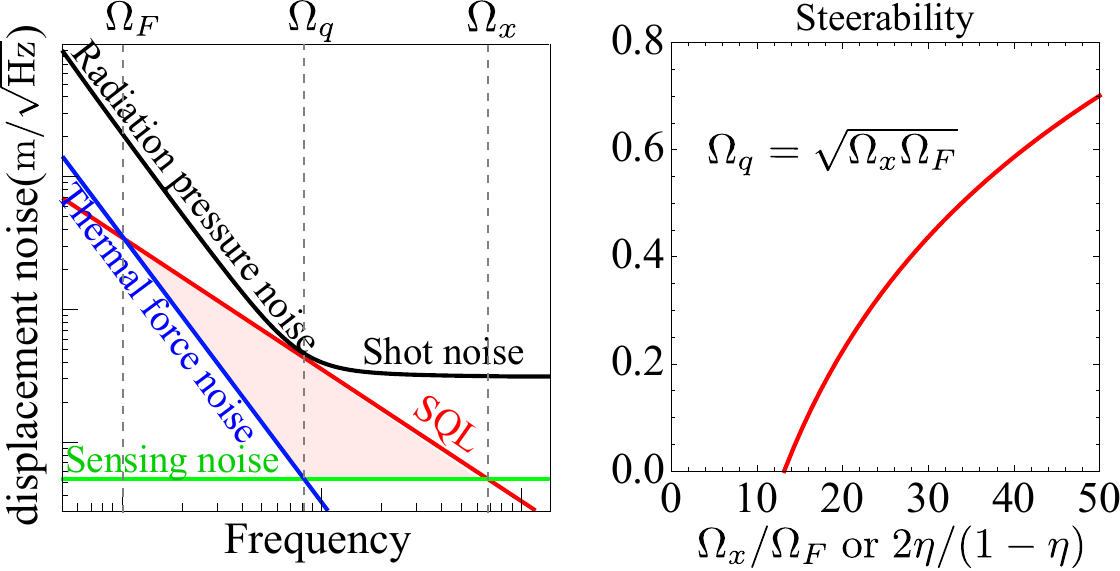}
\caption{(color online) The displacement noise spectrum of an optomechanical device and the characteristic frequencies of the force noise $\Omega_F$, the quantum noise $\Omega_q$, and the sensing noise $\Omega_x$ referring to the Standard Quantum Limit (SQL) (left) and the steerability ${\cal S}_v$ as a function of ratio $\Omega_x/\Omega_F$ while keeping $\Omega_q=\sqrt{\Omega_x\Omega_F}$ (right). Only the ratio between these frequencies determines the steerability, and the absolute values can vary depending on the chosen device.
\label{noise}}
\end{figure}

{\it Verifiable steering and the experimental requirements.---}Not only are the steering and tomography intimately related to each other, but also the tomography is necessary in order to verify the steering in the experiment.
For Gaussian states, the tomography error simply adds on top of the covariance matrix for every conditional state. We therefore define the following figure of merit for
 verifiable quantum-state steering:
\be
{\cal S}_{v}=-\ln\left[ {2}\sqrt{\det[{\bf V}_{s}+{\bf V}_{v}]}/{\hbar}\right]=-\ln(2\zeta_F/\eta).
\ee
We therefore require $\zeta_F<\eta/2$ for verifing quantum-state steering. To illustrate what this condition implies in terms of the displacement sensitivity of such an optomechanical device, in the left panel of Fig.\,\ref{noise}, we compare various displacement noises with respect to the Standard Quantum Limit (SQL)---$S_{x}^{\rm SQL}=2\hbar/(m\Omega^2)$\,\cite{BK92}. We introduce three
characteristic frequencies of these noises where their displacement spectrums intersect the SQL as benchmark through---$\Omega_F\equiv [2m\kappa_m k_B T/(\hbar m)]^{1/2}$, $\Omega_q\equiv [\alpha^2/(\hbar m)]^{1/2}$, and $\Omega_x\equiv \Omega_q[2\eta/(1-\eta)]^{1/2}$. In the right panel of Fig.\,\ref{noise}, we learn that the thermal noise from thermal fluctuation, and the sensing noise from optical loss and quantum inefficiency, need to be at least below the SQL in order to prepare and verify quantum-state steering. Note that the steerability
does not depend on the absolute value of the noise spectrum;
one can therefore have the flexibility to choose the appropriate frequency range to carry out the experiment, depending on the
specific setup.

{\it Acknowledgments.---}We thank  all the members of the AEI-Caltech-MIT MQM group for fruitful discussions.
This research is supported by the Alexander von Humboldt Foundation's Sofja Kovalevskaja Programme, NSF grants
PHY-0956189 and PHY-1068881, the David and Barbara Groce startup fund at Caltech, as well as the Institute for Quantum Information and Matter, a Physics Frontier Center with funding from the National Science Foundation and the Gordon and Betty Moore Foundation.

{\it Notes.---}While preparing this draft, an interesting protocol for realizing steering with pulsed optomechanical devices, instead of using a steady continuous optical field considered here, has been proposed by He and Reid\,\cite{He2012}.


\bibliography{references}


\appendix

\section{Wiener-Hopf method}\label{App}

In this appendix, we show how to derive ${\bf C}^{\rm T}\,{\bf B}^{-1}{\bf C}$ in Eq.\,\eqref{eq_steer}, which is equivalent to
solving an integral equation shown in Eq.\,\eqref{eqint},
with the Wiener-Hopf method. We first show the general formalism and then specialize to the large-bandwidth and strong-measurement limit that we have considered.

\subsection{General formalism}\label{app_GF}

Here we first discuss the general formalism.
 In the continuous-time limit, ${\bf C}$ is a $2\times 2$ matrix
 with elements given by $C_{ij}(t)$ and, similarly,
 the elements of ${\bf B}$ are $B_{ij}(t-t') \,(i,j=1,2)$
 and they only depend on
time difference due to stationarity). We define
${\bf K}={\bf B}^{-1}{\bf C}$ or equivalently, ${\bf B}{\bf K}={\bf C}$, and ${\bf K}$ is a $2\times2$ matrix with the elements satisfying the following integral equations:
\begin{align}\label{app1}
\sum_{k=1}^2 \int_{-\infty}^0 dt' B_{ik}(t-t')K_{kj}(t')=C_{ij}(t)\,,
\end{align}
and therefore, inverting $\bf B$ is equivalent to solving the above integral equation.
Note that the upper limit of the integration is $0$, instead of $+\infty$ in which case it
can be solved simply by using Fourier transform. This arises naturally in
the classical filtering problem with only past data that is available. The procedure for using Wiener-Hopf method to
solve this set of integral equations goes as follows.

{\it Firstly}, we extend the definition of
$K_{ij}(t)$ and $C_{ij}(t)$ to $t>0$ but requiring $K_{ij}(t)=C_{ij}(t)=0$ if $t>0$, namely
\be
K_{ij}(t)\rightarrow K_{ij}(t)\Theta(t),\quad C_{ij}(t)\rightarrow C_{ij}(t)\Theta(t)\,,
\ee
This allows to extend the upper limit of the integral to be $+\infty$ without changing the result.

{\it Secondly}, we apply the Fourier transform
\be\label{app_ft}
\tilde f(\omega)\equiv F[f(t)]= \int_{-\infty}^{+\infty}dt\,e^{i\omega t}f(t)\,,
\ee
of the above equation and obtain
\be\label{app4}
\left [\sum_{k=1}^2 \tilde B_{ik}(\omega)\tilde K_{kj}(\omega)-\tilde C_{ij}(\omega)\right]_-=0\,,
\ee
where $[f(\omega)]_-$ means the part of $\tilde f(\omega)$ that is analytical (no poles)
in the upper-half complex plane by using the following decomposition:
\be
\tilde f(\omega)\equiv [\tilde f(\omega)]_++[\tilde f(\omega)]_-\,,
\ee
and $[f(\omega)]_+$ is the part that is analytical in the lower-half complex plane. From the definition of Fourier transform in Eq.\,\eqref{app_ft}, the inverse Fourier transform of $[\tilde f(\omega)]_-$ vanishes for $t>0$ from the residue theorem, namely
\be
F^{-1}\Big[[\tilde f(\omega)]_-\Big]=f(t)\Theta(t)=0\,,\quad {\forall}\; t>0\,.
\ee

Let us focus on the equations associated with the first column of $\tilde C$, i.e., $j=1$ (the situation for $j=2$ will be similar), and we
rewrite Eq.\,\eqref{app4} explicitly in terms of their components:
\begin{align}\label{app2}
\left[\tilde B_{11}(\omega)\tilde K_{11}(\omega)+\tilde B_{12}(\omega)\tilde K_{21}(\omega)-\tilde C_{11}(\omega)\right]_-&=0\,,\\\label{app3}
\left[\tilde B_{21}(\omega)\tilde K_{11}(\omega)+\tilde B_{22}(\omega)\tilde K_{21}(\omega)-\tilde C_{21}(\omega)\right]_-&=0\,.
\end{align}

{\it Thirdly}, if $B_{11}(t)=B_{11}(-t)$, we can factorize $\tilde B_{11}(\omega)$ as
\be\label{app_B11}
\tilde B_{11}(\omega)=\tilde \varphi_+(\omega)\tilde \varphi_-(\omega)
\ee
which is the Fourier counterpart of the Cholesky decomposition in  time domain. We now can express $\tilde K_{11}(\omega)$ in terms
of $\tilde K_{21}(\omega)$ in Eq.\,\eqref{app2}. We use the fact that
\be\label{app_5}
[\tilde f(\omega)]_-=0\quad \Longrightarrow \quad [\tilde f(\omega)\tilde g_+(\omega)]_-=0\,,\; {\forall \tilde g}
\ee
Multiplying Eq.\,\eqref{app2} by $\varphi_+^{-1}(\omega)$, we get
\be\label{app_K11}
\tilde K_{11}=\frac{1}{\tilde \varphi_-}\left[\frac{\tilde C_{11}}{\tilde \varphi_+}-\frac{\tilde B_{12}\tilde K_{21}}{\tilde \varphi_+}\right]_-\,.
\ee
Plugging it into Eq.\,\eqref{app3}, we obtain
\begin{align}\nonumber
\Bigg\{\left[\tilde B_{22}-\frac{\tilde B_{21}\tilde B_{12}}{\tilde B_{11}}\right]\tilde K_{21} &+\frac{\tilde B_{21}}{\tilde \varphi_-}\left[\frac{\tilde C_{11}}{\tilde\varphi_+}\right]_- \\&+\frac{\tilde B_{21}}{\tilde \varphi_-}\left[\frac{\tilde B_{12}\tilde K_{21}}{\varphi_+}\right]_+-\tilde C_{21} \Bigg\}_-=0
\end{align}
where we have used the fact that:
\be
\frac{\tilde B_{12}\tilde K_{21}}{\tilde \varphi_+}=\left[\frac{\tilde B_{12}\tilde K_{21}}{\tilde \varphi_+}\right]_++\left[\frac{\tilde B_{12}\tilde K_{21}}{\tilde \varphi_+}\right]_-\,.
\ee

Again, if $\tilde B_{22}-\tilde B_{21}\tilde B_{12}/\tilde B_{11}$ is an even function of time---due to stationarity and time-reversal symmetry. We can make a similar factorization to the one shown in Eq.\,\eqref{app_B11}:
\be\label{app6}
\tilde B_{22}(\omega)-\frac{\tilde B_{21}(\omega)\tilde B_{12}(\omega)}{\tilde B_{11}(\omega)}= \tilde \psi_+(\omega)\tilde \psi_-(\omega)\,.
\ee
The same as the approach for deriving Eq.\,\eqref{app_K11} by using the fact shown in Eq.\,\eqref{app_5}, we obtain
\be\label{app_K21}
\tilde K_{21}=\frac{1}{\tilde \psi_-}\left[\frac{1}{\tilde \psi_+}\left(\tilde C_{21}-\frac{\tilde B_{21}}{\tilde \varphi_-}\left[\frac{\tilde C_{11}}{\tilde\varphi_+}\right]_--\frac{\tilde B_{21}}{\tilde \varphi_-}\left[\frac{\tilde B_{12}\tilde K_{21}}{\tilde \varphi_+}\right]_+\right)\right]_-\,.
\ee
Note that in the above equation, $\tilde K_{21}$ appears in both
sides of the equation, which might seems to be difficult to solve.
Actually, since $\tilde K_{21}$ is analytical in the upper half complex plane, $[\tilde B_{12}\tilde K_{21}/\tilde \varphi_+]_+$ only depends on the value of $\tilde K_{21}$ at the poles of $\tilde B_{12}/\tilde \varphi_+$ in the upper-half complex plane. One only need to solve a set of simple algebra equations by evaluating
the above equation on these poles.

\subsection{Large-bandwidth and strong-measurement limit}\label{app_LB}

In this section, we consider the case of the large-bandwidth and strong-measurement limit, and, for the oscillator,
we have
\begin{align}
\hat x(t)&=\int_{-\infty}^{t}{\rm d}t'\,G_x(t-t')[-\alpha \,\hat a_1(t') + \hat F_{\rm th}(t')]\,,\\
\hat p(t)&=m\int_{-\infty}^{t}{\rm d}t'\,\dot G_x(t-t')[-\alpha \, \hat a_1(t') + \hat F_{\rm th}(t')]\,,
\end{align}
and, for the output optical field, we have
\begin{align}
\hat y_1(t)&=\sqrt{\eta}\,\hat a_1(t)+\sqrt{1-\eta}\,\hat n_1(t)\,,\\\label{eq_y2}
\hat y_2(t)&=\sqrt{\eta}\,\left[\hat a_2(t)+({\alpha}/{\hbar})\,\hat x(t)\right]+\sqrt{1-\eta}\,\hat n_2(t)\,.
\end{align}
Here
\be
G_x(t)=\frac{1}{m\omega_m}e^{-\kappa_m t/2}\sin\omega_m t
\ee
is the Green's function of the mechanical oscillator, and in the strong-measurement limit---the frequency at which we carry out the measurement is much higher than that of the mechanical frequency, the oscillator can be treated as a free mass and $
G_x(t)|_{\rm free\;mass}={t}/{m}.$

By using the fact that
\be
\langle \hat a_{j}(t)\hat a_{k}(t')\rangle_{\rm sym}=\langle \hat n_{j}(t)\hat n_{k}(t')\rangle_{\rm sym}=\frac{1}{2}\delta_{jk}\delta(t-t')
\ee
for $j, k =1, 2$, and
\be
\langle \hat F_{\rm th}(t)\hat F_{\rm th}(t')\rangle_{\rm sym}=2m\kappa_m k_B T\delta(t-t')\,,
\ee
we obtains the elements for covariance matrix $\bf B$ of $(\hat y_1,\,\hat y_2)$ in the frequency domain (spectral density):
\begin{align}
\tilde B_{11}(\omega)&=1\,,\\
 \tilde B_{12}(\omega)&=-\eta \frac{\alpha^2}{\hbar}\tilde G_x^*(\omega)\,,\\
\tilde B_{21}(\omega)&=-\eta \frac{\alpha^2}{\hbar}\tilde G_x(\omega)\,,\\
\tilde B_{22}(\omega)&= 1+\eta\frac{\alpha^2}{\hbar^2} \tilde S_{xx}(\omega)\,,
\end{align}
and the correlation between $(\hat y_1,\,\hat y_2)$ and $(\hat x(0)\,,\hat p(0))$:
\begin{align}
\tilde C_{11}(\omega)&=-\sqrt{\eta}\,\alpha \tilde G_x^*(\omega)\,,\\
\tilde C_{12}(\omega)&=-i\,m\Omega\sqrt{\eta}\alpha \tilde G_x^*(\omega)\,,\\
\tilde C_{21}(\omega)&=\sqrt{\eta}\,\frac{\alpha}{\hbar} \tilde S_{xx}(\omega)\,,\\
\tilde C_{22}(\omega)&=i\,m\Omega\sqrt{\eta}\frac{\alpha}{\hbar} \tilde S_{xx}(\omega)\,,
\end{align}
where
\be
\tilde S_{xx}(\omega)\equiv|\tilde G_x(\omega)|^2(\alpha^2+4m\kappa_m k_B T)\,,
\ee
and the Fourier transform for the mechanical Green's function is given by [strong-measurement limit is taken by setting $\gamma_m,\,\omega_m\rightarrow0 $]:
\be
\tilde G_x(\omega)=\frac{-1}{m(\omega^2-\omega_m^2+i\kappa_m\omega)}.
\ee

In this case, we can easily carry out the factorization. For $\tilde \varphi_{\pm}(\omega)$ [cf. Eq.\,\eqref{app_B11}], we have
\be
\tilde \varphi_+(\omega)=\tilde \varphi_-(\omega)=1\,,
\ee
For $\tilde \psi_{\pm}(\omega)$ [cf. Eq.\,\eqref{app6}], we have
\begin{align}\nonumber
\tilde \psi_+(\omega)\tilde \psi_-(\omega)&=1+\eta\frac{\alpha^2}{\hbar^2}|\tilde G_x^2(\omega)|^2\left[(1-\eta)\alpha^2 + 4m\kappa_m k_B T\right]\\\label{app7}
&=\frac{\omega^4+(\kappa_m^2-2\omega_m^2)\omega^2+\omega_m^4+2(\alpha^2/\hbar m)^2\zeta_F^2}{\omega^4+(\kappa_m^2-2\omega_m^2)\omega^2+\omega_m^4}\,.
\end{align}
This leads to
\be
\tilde\psi_+(\omega)=\tilde \psi_-^*(\omega)=\frac{(\omega-\omega_1)(\omega-\omega_2)}
{(\omega-\omega'_1)(\omega-\omega'_2)},
\ee
where $\omega_j$ and $\omega_j'$ $(j=1,2)$ are the roots of the numerator and denominator of Eq.\,\eqref{app7} in the upper-half complex plane, respectively. Given the expression for $\tilde \varphi_{\pm}$ and $\tilde \psi_{\pm}$, we can solve $\tilde K_{ij}$ by using Eqs\,\eqref{app_K11}, and Eq.\,\eqref{app_K21}. This  in turn allows us to obtain ${\bf V}_s = {\bf A}-{\bf C}^{\rm T}{\bf K}$ and in the time domain, it reads:
\be
({\bf V}_s)_{ij}={\bf A}_{ij}- \sum_{k}\int_{-\infty}^0{\rm d}t' \, C_{ki}(t')K_{kj}(t')\,,
\ee
from which we obtain Eq.\,\eqref{eq_Vs}.

\end{document}